\documentstyle[amssymb,aps,prb,twocolumn,psfig]{revtex}

\draft

\begin{document}

\twocolumn[\hsize\textwidth\columnwidth\hsize\csname@twocolumnfalse\endcsname

\title{Double Exchange in Electron Doped Ca$_{1-x}$Y$_{x}$MnO$_{3}$ Manganites }
\author{H. Aliaga, M. T. Causa, H. Salva, M. Tovar, A. Butera and B. Alascio.}
\address{Centro At\'omico Bariloche and Instituto Balseiro}
\address{Comisi\'on Nacional de Energ{\'\i}a At\'omica and Universidad Nacional de Cuyo.}
\address{8400 San Carlos de Bariloche, Argentina.}
\author{D. Vega, G. Polla, G. Leyva, and P. Konig}
\address{Centro At\'omico Constituyentes.}
\address{Comisi\'on Nacional de Energ{\'\i}a At\'omica}
\address{1650 San Mart{\'\i}n, Buenos Aires, Argentina.}

\date{\today}

\maketitle

\begin{abstract}

We have studied structural, magnetic and transport properties as a function
of temperature and magnetic field in the electron doped manganite
Y$_{x}$Ca$_{1-x}$MnO$_{3}$,
for 0$\leq x\leq $0.25.  We found that in the paramagnetic regime, the magnetic
susceptibility, $\chi$, deviates substantially from a Curie-Weiss law for $x>$0.  With a
simple model where
antiferromagnetic (AF) superexchange and ferromagnetic (FM) double exchange (DE) compete,
we fit the experimental $\chi(x, T)$ obtaining parameter values which indicate that the FM-DE
interaction is about twice as intense as the AF interaction.  In the ordered phase, the
$H$-dependence of the magnetization $M(x,T)$ is explained in terms of magnetic polarons.  We
propose that the displacement of the $e_{g}$ electrons from one magnetic sublattice to other
(in the G-type AF background) causes the alignement of the polaron with H.  Signatures
of polaronic behavior were also found in the $x$ and T dependence of the electric resistivity.  
\end{abstract}

\pacs{PACS numbers: 75.30.Vn, 75.10.-b, 75.40.Mg }
]

\section{Introduction}

The existence of magnetoresistance (MR) in Mn oxides with perovskite
structure was discovered at the very early stage of the study of the
transition metal oxides\cite{tok2}. The interest in these compounds has
revived recently due to the discovery of very large MR ($\sim 10^{6}\%$)
which led to call this phenomenon colossal magnetoresistance (CMR). At the
same time a magnetic field induced insulator-metal and structural transition%
\cite{TT} was also discovered. Most studies were devoted to the perovskite
compounds R$_{x}$A$_{1-x}$MnO$_{3}$ (R=trivalent rare earth and A=divalent
alkaline earth) producing $x$ Mn$^{3+}$ and (1-$x$) Mn$^{4+}$ ions,
respectively. Ferromagnetic (FM) double-exchange (DE) interaction between
localized $t_{2g}$ Mn electron configuration, mediated by itinerant
spin-polarized $e_{g}$ electrons, is in the base of CMR. The concentration $x
$ is not the only parameter to be taken into account in CMR materials.
Chemical parameters such as the average R-A cationic radius and the cationic
size mismatch quantified by the variance $\sigma ^{2}$ of the ionic radii%
\cite{Attfield,Rav} are also relevant. The phase diagrams for R$_{x}$A$_{1-x}
$MnO$_{3}$ families were found to be non-symmetric and hole or electron
doping cause disimilar effects. Theoretical studies\cite{MD} suggested at
least three possible scenarios to understand the low electron doping region:
i) Canting of the magnetic structure. ii) Phase separation into
ferromagnetic and antiferromagnetic phases with charge separation. iii)
Ferromagnetic polarons. Therefore, experiments on families of compounds,
covering different ranges of physical and chemical parameters can be used to
clarify this issue. The most studied series are those where A$=$Ca. In Ref.
6 an analysis of the saturation magnetization (M) of Ca$_{1-x}$R$_{x}$MnO$%
_{3}$ was performed for different R ions. They found, in all cases, an
initial rise of M with $x$ followed by a sudden drop associated to charge
ordering (CO). The value of $x$ where the CO phase appears depends on the size
of R: a higher electron concentration is necessary to establish the AF-CO
state as the ionic radius of R decreases. In this paper we study structural,
transport, and magnetic properties of Ca$_{1-x}$Y$_{x}$MnO$_{3}$. Trivalent Y
is one of the smallest ions synthesizing in the perovskite structure. Then
the charge ordering phase is expected for higher $x$ in comparison with the
compounds studied in Ref. 6 and a more detailed analysis of the initial
magnetic phase can be made. The Y series synthesizes\cite{CAC} with orthorrombic
structure for 0$<x<$0.8. Magnetoresistance effects were
reported\cite{Icm2000} for 0.05$\leq x\leq $0.15, indicating the presence
of DE interaction. The non magnetic character of Y ions makes the series Ca$%
_{1-x}$Y$_{x}$MnO$_{3}$ an excellent system to study the evolution of the
magnetism of Mn ions, without interference from other magnetic species. This
is specially important in the paramagnetic regime. We present here magnetic
measurements in the region 0$\leq x\leq $0.25 and compare with the
resistivity. We analyse these measurements in the PM and ordered phases to
show that they can be understood in terms of small magnetic polarons\cite
{Bati} arising from the DE mechanism between Mn$^{4+}$ and Mn$^{3+}$. We
will show that this model explains also experimental results reported in the
literature\cite{Rav,Maignan,Troyanchuk,Neumeyer,Goodwin,Chiba} for Ca$_{1-x}$%
R$_{x}$MnO$_{3}$ with $x$ values covering the electron doped region and R
different from Y.

\section{Experimental}

Ceramic polycrystalline samples of Ca$_{1-x}$Y$_{x}$MnO$_{3}$ were prepared
by solid state reaction methods\cite{CAC,Icm2000}. Room temperature x-rays
diffractograms show that CaMnO$_{3}$ crystallizes in an orthorhombic $Pnma$
cell with parameters $a$=5.284(5)\AA , $b$=7.453(5)\AA , and $c$=5.266(5)\AA . For
0$\leq x\leq $0.25 all the samples crystallize in single orthorrombic
phase O (with $c<b/\sqrt{\text{2}}$). In Fig. 1 we show the variation of
the cell parameters in the range studied here. The structural distortion
increases with $x$ and can be evaluated by the orthorrombic strain: s$=%
2(a-c)/(a+c)$, shown in the inset of Fig. 1. The tilt and rotation angles,
produced respectively by apical and planar oxygens shift, remains
approximately constant. A study of the cell parameters evolution with T, for 
$x$=0.07, shows no structural transition in the range 15-300K. Electrical
resistivity was measured with the 4-probe method using a current source of 10%
$-$100$\mu $A. The dc-magnetization $M$\ was measured with a SQUID
magnetometer between 5K and 300K for $H\leq $50kOe and with a Faraday
Balance magnetometer between 300K and 1000K with $H\leq $10kOe.

\begin{figure}[t]
\centerline{\psfig{figure=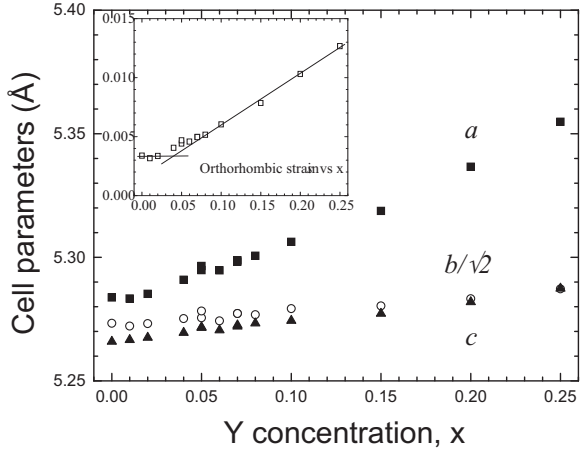,height=6cm,angle=0}}
\medskip
\caption{Room temperature cell parameters $a$, $b$, and $c$ as a function of $x$. In
the inset, the orthorrombic strain $s$ vs. $x$ is shown.}
\end{figure}

\section{Results}

In Fig. 2 we show $H/M$ vs. $T$ measured for the samples $x=$0, 0.05, 0.07,
and 0.10. In all cases $M$ is a linear function of $H$ for $T\geq $140K. A
Van Vleck contribution is expected for Mn$^{4+}$ ions\cite{Huber}. In our
measurements this contribution was subtracted in the paramagnetic phase,
proportionally to (1-$x$). The CaMnO$_{3}$ data were fitted with a Curie Weiss
law, $\chi =C/(T-\Theta )$, where $C$ is the Curie constant and $\Theta $ is
the Curie-Weiss temperature, obtaining $C=$(1.95$\pm $0.05) emu-K/mol (near
the expected Mn$^{4+}$ value 1.875) and $\Theta \approx $-400K$.$ This value
points to a strong antiferromagnetic (AF) superexchange interaction between
Mn$^{4+}$ ions. In this case $M$ remains linear with $H$ down to $T_{N}$.
Using neutron diffraction, Wollan and Kohler\cite{WoKo} found G-type AF
ordering below $T_{N}\sim $123K. The ratio $\Theta /T_{N}>$3 is an
indication of the importance of second neighbors interaction in the
perovskite structure\cite{Huber}. Further magnetic measurements have shown
the existence of a weak ferromagnetic moment\cite{WF} below $T_{N}$ of $%
M_{WF}\sim $0.03$\mu _{B}$/Mn ion, associated with Dzialoshinsky-Moriya (DM)
interaction. The hysteresis loop for this weak ferromagnetic component shows
a small coercive field (H$_{c}$=0.12T at 5K). Since the reversal of $M_{WF}$
implies reversal of the full lattice magnetizations, the H$_{c}$ observed
suggest a small anisotropy energy.

\begin{figure}[t]
\centerline{\psfig{figure=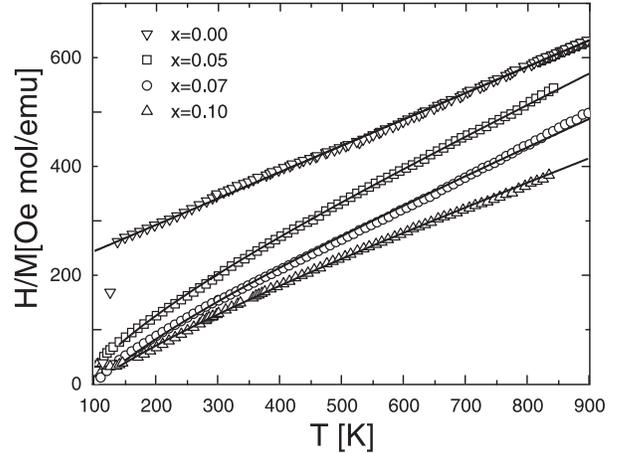,height=6cm,angle=0}}
\medskip
\caption{Measured $\chi ^{-1}$(T) curves (symbols) with H=10kOe and
calculated (lines) for the samples $x=$0, 0.05, 0.07 and 0.10. Notice the
deviation from a Curie-Weiss law below $\simeq 450K$.}
\end{figure}

At high temperatures, the samples with $x>$0 also follow a Curie-Weiss law.
However, for $T\lesssim $450K $H/M$ deviates substantially from this
behaviour. In Fig. 3(a) it is shown the dependence of $\Theta $ with the
doping $x$, as obtained from high T fits. Small $Y$ doping causes large
changes in the $\Theta $ values, indicating an evolution from a strong AF
for $x=$0 ($\Theta \approx $-400K) to a FM one for $x\approx $0.10 ($\Theta
\approx +$80K). In Fig. 4 we show $M$ vs. $T$ curves measured with $H=$5kOe,
for selected samples. Below $T\sim $125K a FM component that increases with 
$x$ up to $x=$0.10 was measured. A magnetic transition temperature, $T_{mo}$,
was defined as the temperature with maximum slope of $M$($T$). A detailed
dependence of$\ T_{mo}$ vs. $x$ is shown in Fig. 3(b).

In Fig. 5 we show $M$ vs. $H$ at $T=$ 5K. At low fields ($H<$5kOe), $M(H)$
varies rapidly with $H$ and, for $H$=0 there is only a small remanent
magnetization. For $|H|\gtrsim $30kOe, $M(H)$ can be approximated by: $%
M=M_{0}+\chi _{diff}$ $H$, where $M_{0}$ is the ferromagnetic contribution
to $M$ and $\chi _{diff}$ is the high field differential susceptibility$.$
In the case of $x$=0, $M_{0}$=$M_{WF}$ is the DM contribution and $\chi _{diff}
$ is the AF susceptibility. Measurements between 5K and $T_{N}$ shows that $%
\chi _{diff}(T)$ remains almost constant, after subtraction of a small
Curie-type contribution observed below 25K (equivalent to about 0.3\% of Mn$%
^{4+}$ free ions).

\begin{figure}[t]
\centerline{\psfig{figure=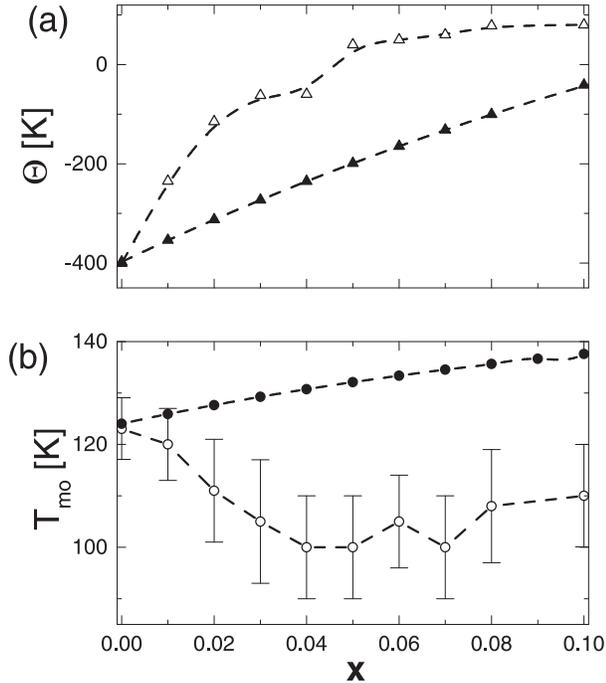,height=9cm,angle=0}}
\medskip
\caption{(a) Curie-Weiss temperatures, $\Theta $ vs. $x$. Open symbols are
determined from linear fits above 500K. Solid symbols are derived from Eq.
7. (b) Measured (open) and calculated (solid) magnetic ordering temperature, 
$T_{mo}$ vs. $x$.}
\end{figure}

\begin{figure}[t!]
\centerline{\psfig{figure=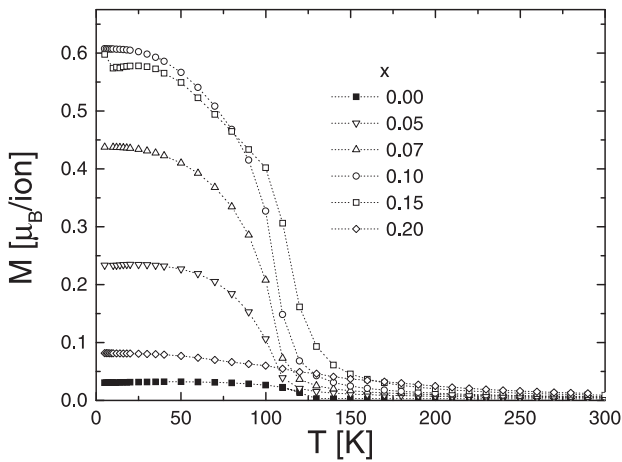,height=6cm,angle=0}}
\medskip
\caption{M vs. T curves with H=5kOe. Lines are guides to the eye.}
\end{figure}

For $x>$0 a similar behaviour was observed at all temperatures measured,
with an important increase of both $M_{0}$ and $\chi _{diff}.$ We have
measured in detail the temperature dependence of $M$ vs. $H$ for $x$=0.1. The
resulting $M_{0}(T)$ and $\chi _{diff}(T)$ are compared in Fig. 6 with the
case $x$=0. The susceptibility $\chi _{diff}(T)$ shows a maximum in
coincidence with $T_{mo}$. This behaviour resembles that of an
antiferromagnet, although with a much large susceptibility than that of the
undoped CaMnO$_{3}$.

In Fig. 7(a) we plot $M_{0}$ vs. $x$ at 5K. Notice that $M_{0}$ is
much smaller than the expected value for full FM alignment ($M_{S}\sim $3$%
\mu _{B}$/Mn ion). In the region 0$\leq x\leq $0.03, $M_{0}(x)$ increases
almost linearly, with an initial slope of $\sim $0.85$\mu _{B}$/Mn ion. For $x$%
$\gtrsim $0.03, $M_{0}$ increases more rapidly and reaches a maximum ratio $%
M_{0}/x\approx $ 7.1$\mu _{B}$/Mn ion for $x\sim $ 0.07-0.1. For $x>$0.15, 
$M_{0}$ decreases sharply.

\begin{figure}[t!]
\centerline{\psfig{figure=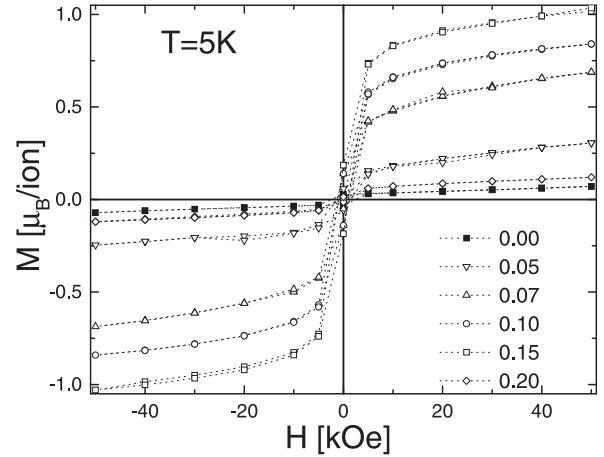,height=6cm,angle=0}}
\medskip
\caption{M vs. H curves at 5K. Full alignment ($M_{0}\sim 3\mu _{B}$) is not
reached in any case. Dotted lines are guides to the eye.}
\end{figure}

In Fig. 8(a) we present results for the resistivity in the range 4-300K. The
sample $x=$0 is the most resistive one, and a strong decrease in $\rho $ is
induced with small electronic doping $x$. Small features are
found in $\rho(T)$ near $T_{mo}$. These curves were fitted in the range
100-300K assuming a polaronic behaviour $\rho (T)$ $\alpha $ $T\exp [-\Delta
/k_{B}T]$, showing double exchange effects and activation energies
of 1160, 390, 420, 490, 460, and 455K for the samples $x=$0, 0.01, 0.03, 0.05,
0.08 and 0.10, respectively. The dependence of $\rho(T) $ in the range 4K$\leq
T\leq $100K, is more complicated and shows a decrease of dln$\rho $/d$T$ at
low $T.$ This behaviour indicates that the material is not an insulator.

In Fig. 7(b) we present the dependence of the electric resistivity $\rho $%
(5K). The resistivity decreases orders of magnitud between $x=$0 and $x=$%
0.05, remains almost constant for 0.05$\leq x\leq $0.15, and then
increases again for 0.15$\leq x\leq $0.25. Comparing Figs. 7(a) and 7(b)
in the range 0$\leq x\leq $0.15, we see that the greater the concentration
of carriers, the greater the conductivity and the magnetic moment. Fig. 8(b) shows
the linear dependence found for the conductivity, $\sigma $ vs. $x$.
This result is in qualitatively agreement\cite{Zener} with the DE mechanism. In
the region 0.15$\leq x\leq $0.25, the resistivity increases and the
magnetic moment decreases, the DE mechanism is broken, leading to a low
magnetic moment-insulator state.

\begin{figure}[t]
\centerline{\psfig{figure=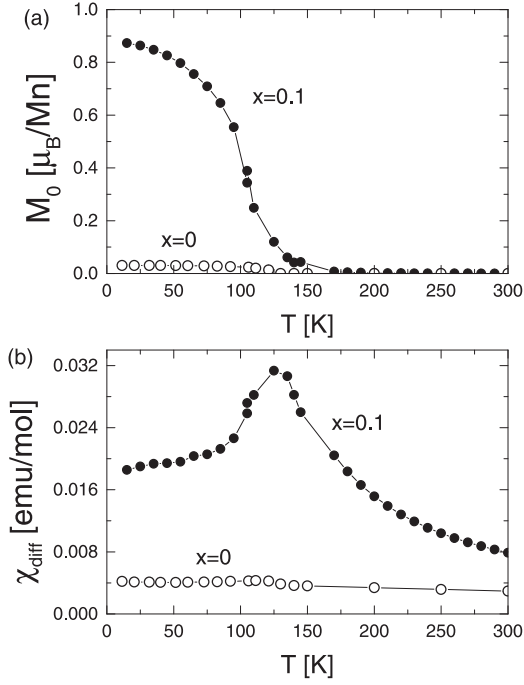,height=9cm,angle=0}}
\medskip
\caption{T dependence of: (a) $M_{0}$  and (b) $\chi _{diff}$,  for $x$=0 and
0.1, obtained from M vs. H in the high field linear regime.}
\end{figure}

\begin{figure}[t]
\centerline{\psfig{figure=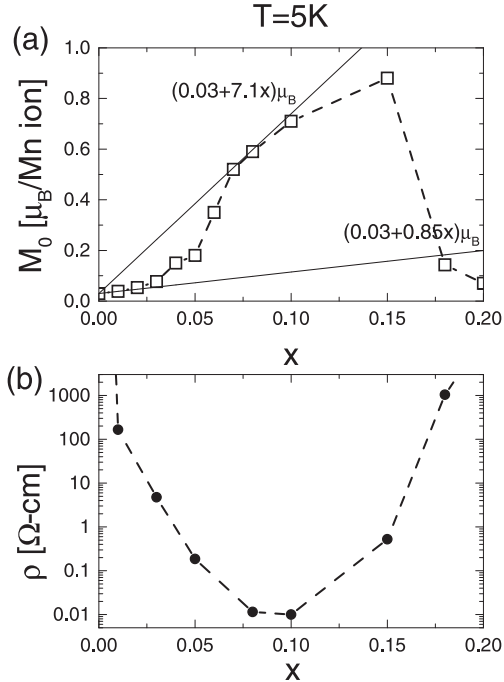,height=9cm,angle=0}}
\medskip
\caption{(a) High field extrapolated magnetization $M_{0}$ vs. $x$.
Continous straight lines indicates the initial and maximum slopes. (b) The
zero field resistivity $\rho $ vs. $x$. Notice the strong correlations
between magnetism and resistivity. Dotted lines are guides to the eye.}
\end{figure}

\begin{figure}[t]
\centerline{\psfig{figure=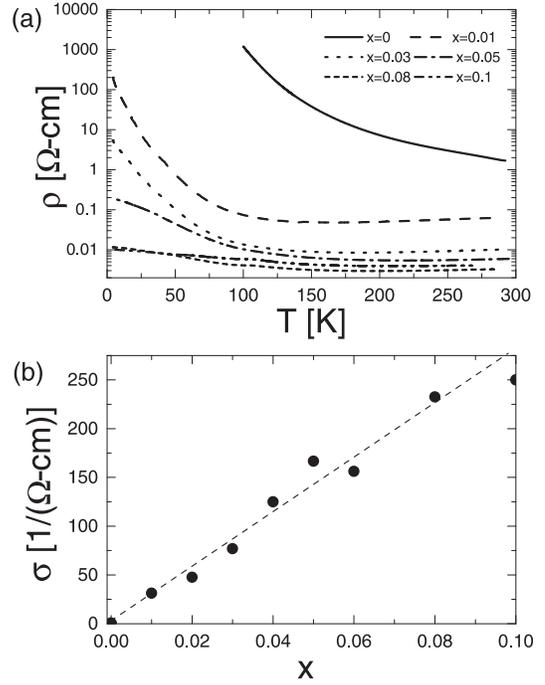,height=9cm,angle=0}}
\medskip
\caption{(a) $\rho $ vs. $T$ for the samples $x=$0, 0.01, 0.03, 0.05, 0.08,
0.1, and 0.15. (b) $\sigma $ vs. $x$ at 300K.}
\end{figure}

\section{Discusion}

When Y$^{3+}$ substitutes for Ca$^{2+}$ in CaMnO$_{3}$, $e_{g}$ electrons
are introduced in the system. These electrons polarize the otherwise G-type
AF background, forming magnetic polarons\cite{Bati}. In the very dilute
region, we expect that Jahn-Teller distortions surrounding each Mn$^{3+}$
will tend to localize the $e_{g}$ electrons, forming a combined
lattice-magnetic polaron. In this regime each polaron carries an extra
magnetic moment due to the added $e_{g}$ electron ($S=$1/2). Within this
model the large variation of $M$ with $H$ for $|H|\leq $10kOe (see Fig. 5)
at low temperatures is mainly due to the alignment of these small polarons\cite{Neumeyer}.
With zero applied field the Mn$^{3+}$ spins, with a magnetic of 4$\mu _{B}$,
are oriented in random up and down directions in the lattice and the total
average moment is null. With a nonzero field, polarons may easily reorient
themselves parallel to the field, through the hopping of the $e_{g}$
electron to a neighboring site using the DE mechanism. Fig. 9(a) sketches the
suggested behaviour in the presence of a field H. This process results in a
linear dependence: $M_{0}/x\sim $1$\mu _{B}$/Mn ion.

Neutron diffraction experiments\cite{Woodward} show that, in the case of 
Ca$_{1-x}$Bi$_{x}$MnO$_{3}$, an esentially G-type AF background is preserved up to
$x\sim$0.1. We assume a similar scenario for Ca$_{1-x}$Y$_{x}$MnO$_{3}$ since the 
lattice presents the same $Pnma$ structure for the whole composition range 
of our samples.
Small distortions of the perfect AF alignment are produced by the DE interactions, 
allowing the excitations of the $e_{g}$ electrons from the localized Jahn-Teller 
states to a narrow conduction band, improving in this way, the electrical 
conductivity.

\begin{figure}[t!]
\centerline{\psfig{figure=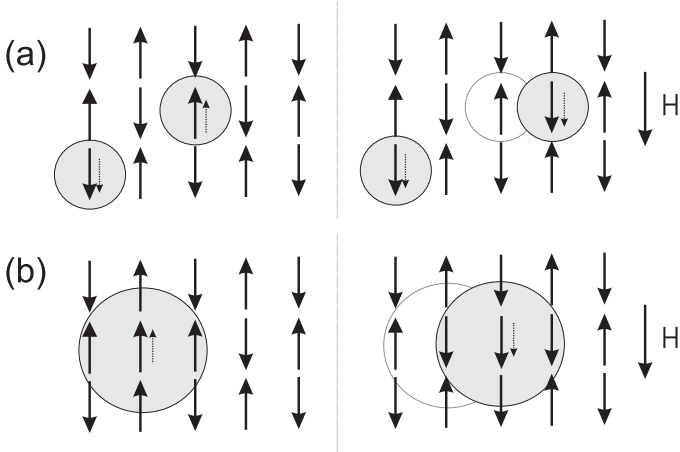,height=6cm,angle=0}}
\medskip
\caption{(a) Each $e_{g}$ electron added to the system couples
ferromagnetically with the localized spin $t_{2g}$ forming single-site
polarons for small $x$. (b) For larger $x$, the size of the polarons 
increases, involving the neighborings Mn spins. spin-flip processes. The 
figure illustrates the proposed mechanism of magnetization reversal, with 
an applied field, throug the hopping of $e_{g}$ electrons.}
\end{figure}

For 0.03$\lesssim x\lesssim $0.07, the $M_{0}$ vs. $x$ curve displays a
pronounced concavity. For $x\geq $0.07, we find again a linear regime with 
$M_{0}/x\sim $7.1$\mu _{B}$/Mn-ion, indicating that the polarons carry a
larger magnetic moment. 

Changes in the $x$ dependence of $M_{0}$ may be due
to the occupancy of states above the bottom of the band\cite{Brink}, to
effective interaction between polarons, or to variations of the
cristallographic structure, that may favour different magnetic arrangements. 
The x-ray diffractograms show that all the
samples may be indexed in the same orthorrombic O-phase at room temperature
and no structural transitions has been observed between 15K and 300K for 
$x$=0.07. Thus we have discarded the appearence of magnetic phases (A- or C-like) 
other than
the distorted G-type, since these phases have been observed for higher
electron doping\cite{Kajimoto}, and associated with structural transitions%
\cite{Mahendiran} as in Ca$_{0.85}$Sm$_{0.15}$MnO$_{3}$. We can
interpret the results in the region 0.07$\lesssim x\lesssim $0.15 with the
following model: each Mn$^{3+}$ couples
ferromagnetically with the Mn$^{4+}$ neighbors, after a spin flip process in
the G-type AF lattice, see Fig. 9(b). This ferromagnetic coupling is
favoured by the kinetic energy gain due to the hopping of the $e_{g}$
electron into the neighboring localized spins, forming a larger magnetic
polaron, comprising the Mn$^{3+}$ site and its nearest neighbors. This
process is made possible by the larger orthorrombic distortion of the whole
lattice (Fig. 1) that provides distorted Mn$^{4+}$ sites where the $e_{g}$ electrons
may jump more easily. With zero applied field, these polarons, are randomly oriented. 
When a magnetic field $H$ is applied, the $e_{g}$
electrons of the misaligned Mn$^{3+}$ spins hop to a Mn$^{4+}$ neighbor,
which has its neighbors oriented in the direction of H. After two
spin-flip processes, see Fig. 9(b), the displaced polarons are aligned with the
field and $M_{0}/x=$7$\mu _{B}$/Mn-ion is expected. The linear increase of 
$M(H)$ observed at high fields for all concentrations, could be interpreted as due 
to the contribution of the AF background. However, notice that $\chi_{diff}$ is 
much larger than the AF susceptibility measured for $x=$0 (see Fig. 6). This fact 
suggest that the application of high magnetic fields enhances the response of the 
Mn$^{4+}$ ions beyond the boundaries of the $H=$0 polarons.

In the paramagnetic phase, the most peculiar feature observed is the
deviation from the Curie-Weiss law, presenting a negative curvature of the $%
\chi ^{-1}(T)$ curves (Fig. 2). In order to describe this behaviour in
simple terms, we consider the system as a mixture of Mn$^{4+}$ and Mn$^{3+}$
ions. Taking into account that the ordered state for $x=$0 is a G-type
antiferromagnet we divide the system into two interpenetrated sublattices $a$
and $b$, where Mn$^{4+}$ and Mn$^{3+}$ are randomly distributed with
concentrations (1-$x$) and $x$, respectively. In a mean field approximation we
consider superexchange interactions with first and second neighbors, between
Mn$^{4+}$-Mn$^{4+}$, Mn$^{3+}$-Mn$^{3+}$ and DE between Mn$^{4+}$-Mn$^{3+}$
pairs. The equations satisfied by the sublattice magnetizations are:

\begin{eqnarray}
M_{4}^{a}T &=&C_{4}[H+\gamma _{44}(1-x)M_{4}^{b}+\gamma _{44}^{\prime
}(1-x)M_{4}^{a}  \nonumber \\
&&+\gamma _{43}xM_{3}^{b}+\gamma _{43}^{^{\prime }}xM_{3}^{a}]
\end{eqnarray}
\begin{eqnarray}
M_{4}^{b}T &=&C_{4}[H+\gamma _{44}(1-x)M_{4}^{a}+\gamma _{44}^{\prime
}(1-x)M_{4}^{b}  \nonumber \\
&&+\gamma _{43}xM_{3}^{a}+\gamma _{43}^{^{\prime }}xM_{3}^{b}]
\end{eqnarray}
\begin{eqnarray}
M_{3}^{a}T &=&C_{3}[H+\gamma _{33}xM_{3}^{b}+\gamma _{33}^{\prime
}xM_{3}^{a}+\gamma _{43}(1-x)M_{4}^{b}  \nonumber \\
&&+\gamma _{43}^{^{\prime }}(1-x)M_{4}^{a}]
\end{eqnarray}
\begin{eqnarray}
M_{3}^{b}T &=&C_{3}[H+\gamma _{33}xM_{3}^{a}+\gamma _{33}^{\prime
}xM_{3}^{b}+\gamma _{43}(1-x)M_{4}^{a}  \nonumber \\
&&+\gamma _{43}^{^{\prime }}(1-x)M_{4}^{b}]
\end{eqnarray}
where subindexes $4$ and $3$ indicate $Mn^{4+}$ and $Mn^{3+}$ ions
respectively, superindexes $a$ y $b$ correspond to the two sublattices, and $%
\gamma _{ij}$ are the parameters describing the exchange coupling between $%
M_{i}$ and $M_{j}$. Primed parameter indicate interaction with second
neighbors. The total magnetization is given by:

\begin{equation}
M=\frac{1}{2}[(M_{4}^{a}+M_{4}^{b})(1-x)+(M_{3}^{a}+M_{3}^{b})x]
\end{equation}

In the PM regime we take $M_{4}^{a}=M_{4}^{b}$ and $M_{3}^{a}=M_{3}^{b}.$
The behaviour of the reciprocal susceptibility $\chi ^{-1}(T)\,$is a
hyperbola and only three independent parameters ($\gamma _{44}+\gamma
_{44}^{\prime },$ $\gamma _{33}+\gamma _{33}^{\prime }$ and $\gamma
_{43}+\gamma _{43}^{\prime }$) are necessary to fit the experimental curves,
since first and second neighbors contribution cannot be separated. Best fits
of the experimental data are shown as continuos lines in Fig. 2, where an
excellent agreement is observed. For $x=$0 the solutions of Eqs. 1-4
correspond to a Curie-Weiss behaviour with a characteristic temperature $%
\Theta =C_{4}(\gamma _{44}+\gamma _{44}^{\prime })$. From the fit, we
derived $\gamma _{44}+\gamma _{44}^{\prime }=$ -204 mol/emu. For 0$<x\leq$0.10, 
the contribution to the total magnetization of the $Mn^{3+}$
ions is small compared to the contribution of the $Mn^{4+}$ species. For
this reason we expect great indetermination in $(\gamma _{33}$+$\gamma
_{33}^{\prime })$, if left as a free parameter. Thus we have taken as an
approximate fixed value, $(\gamma _{33}$+$\gamma _{33}^{\prime })=$ +73
mol/emu, as obtained for the Mn$^{3+}$-Mn$^{3+}$ interaction\cite{Huber} in
the pseudo-cubic high temperature phase of LaMnO$_{3}$. Therefore, when
fitting the data for $x>$0, only ($\gamma _{43}$+$\gamma _{43}^{\prime }$)
is kept as an adjustable parameter. We obtained +361, +386 and +310 mol/emu
for $x=$0.05, 0.07, and 0.10, respectively. We have found that these results
are rather insensitive to the value assumed for the Mn$^{3+}$-Mn$^{3+}$
interaction, within the experimental uncertainty ($\sim 10\%$), if we take
-204 mol/emu $\lesssim $($\gamma _{33}$+$\gamma _{33}^{\prime }$)$\lesssim $
73 mol/emu. The ($\gamma _{43}+\gamma _{43}^{\prime }$) parameter is always
positive and aproximately two times larger than $\gamma _{44}+\gamma
_{44}^{\prime }$, denoting strong ferromagnetic coupling between Mn$^{4+}$
and Mn$^{3+}$ pairs. From now on, we use the average value $(\gamma
_{43}+\gamma _{43}^{\prime })=$+350 mol/emu for all $x$.

The high temperature asymptote of $\chi ^{-1}(T)$ has a Curie -Weiss form,
with

\begin{equation}
C=(1-x)C_{4}+xC_{3}
\end{equation}

and

\begin{eqnarray}
\Theta =C^{-1}[(1-x)^{2}C_{4}^{2}(\gamma _{44}+\gamma _{44}^{\prime}) 
\nonumber \\
+2(\gamma _{43}+\gamma _{43}^{\prime
})x(1-x)C_{3}C_{4}+x^{2}C_{3}^{2}(\gamma _{33}+\gamma _{33}^{\prime })]
\end{eqnarray}

In Fig. 3(a) we compare the values of $\Theta $ derived from this equation
with those determined from a linear fit between 500K and 850K. Notice that
the measured $\chi ^{-1}$(T) curves do not completely reach the asymptotic
linear regime in the temperature range of the experiments.

Within the mean field approximation used, it is also possible to express the
solutions of Eqs. (1)-(4) in terms of two separate contributions

\begin{equation}
\chi (T)=\frac{C_{AF}}{(T-\Theta _{AF})}+\frac{C_{FM}}{(T-\Theta _{FM})}
\end{equation}

where, for $x<<$1,

\begin{equation}
C_{AF}=(1-x)C_{4}+xC_{3}(1-\Gamma ^{2})
\end{equation}

\begin{equation}
C_{FM}=xC_{3}\Gamma ^{2}
\end{equation}

with

\begin{equation}
\Gamma =1-(\gamma _{34}+\gamma _{34}^{^{\prime }})/(\gamma _{44}+\gamma
_{44}^{^{\prime }})
\end{equation}

and

\begin{equation}
\Theta _{AF}=-(1-x)C_{4}(\gamma _{44}+\gamma _{44}^{^{\prime }})+xC_{3}\frac{%
(\gamma _{34}+\gamma _{34}^{^{\prime }})^{2}}{(\gamma _{44}+\gamma
_{44}^{^{\prime }})}
\end{equation}

\begin{equation}
\Theta _{FM}=xC_{3}\left[ (\gamma _{33}+\gamma _{33}^{^{\prime }})-\frac{%
(\gamma _{34}+\gamma _{34}^{^{\prime }})^{2}}{(\gamma _{44}+\gamma
_{44}^{^{\prime }})}\right]
\end{equation}

Using the values of $(\gamma _{ij}+\gamma _{ij}^{^{\prime }})$ previously
obtained, we find $\Theta _{AF}<0$ and $\Theta _{FM}>0$, corresponding to AF
and FM behaviour respectively. 

This picture shows a FM-like component of the
magnetization proportional to $x$ (Eq. 10). In the absence of $Mn^{3+}-Mn^{4+}$
interaction, this contribution corresponds to isolated $Mn^{3+}$ ions. The
effect of FM-DE coupling enhances the magnetic moment of the $Mn^{3+}$ ions (%
$\Gamma >$1), reflecting the magnetic polaron formation. Using the values
derived from the fitting of $\chi (T),$ we obtained an effective moment for the 
polarons, $\mu _{eff}=2.71\mu
_{eff}(Mn^{3+})$. This result suggest that the effective number of Mn$^{4+}$
neighbors involved in the paramagnetic polarons is $z_{eff}\simeq $5.

In order to analyze the magnetic ordering temperature, Eqs. (1-4) must be
solved for $H=0$. The highest temperature that allows nontrivial solution
for $M_{i}$ corresponds to the magnetic transition $T_{mo}$. For $x=$0, AF
ordering is achieved for $T_{mo}=C_{4}$ (-$\gamma _{44}$ + $\gamma
_{44}^{\prime }$). The values $\gamma _{44}=$-134 mol/emu and $\gamma
_{44}^{\prime }=$ -71 mol/emu are derived from the measured\cite{Huber} $%
T_{N}=$123 K. The magnetic state found preserves the AF
alignment between $M_{4}^{a}$ and $M_{4}^{b}$ . The Mn$^{3+}$ ions are
oriented FM with respect to their Mn$^{4+}$ first neighbors, irrespective 
of the value assumed for the Mn$^{3+}$- Mn$^{3+}$
interaction. In this limit the ordering temperature is given by

\begin{equation}
T_{mo}\approx T_{N}(1-x)+(C_{3}C_{4}/T_{N})(\gamma _{43}-\gamma
_{43}^{\prime })^{2}x(1-x)
\end{equation}
where $T_{N}=T_{mo}$ ($x=$0). The first term reflects the dilution of the Mn$%
^{4+}$ lattice and the second the effect of Mn$^{3+}$ - Mn$^{4+}$
interactions. We have varied $\gamma _{43}$ and $\gamma _{43}^{\prime }$ in
order to reproduce the $x$ dependence of $T_{mo}$, shown in Fig. 3(b). For $%
\gamma _{43}$ = 215 mol/emu and $\gamma _{43}^{\prime }=$135 mol/emu, we
found that $T_{mo}$ increases continuously from $x=$0 to $x=$0.1, where a
value of 137 K is reached, reproducing the experimental tendency for $x>$%
0.04. The initial decrease of $T_{mo}$ cannot be included in this
description with a unique set of parameters in Eq. (14).

\section{Conclusions}

We have measured the temperature and magnetic field dependence of structural, 
magnetic and transport properties in the manganites Ca$_{1-x}$Y$_{x}$MnO$%
_{3} $, for 0$\leq x\leq $0.25. Despite of the small cationic size of Y,
and the large mismatch between Y and Ca, we have synthesized single
phase material with orthorhombic structure. Strong correlation between
magnetization and conductivity in the region 0$\leq x\leq $0.15 was found,
allowing to conclude that DE mechanism is present in this compound.

From the $M$ vs. $H$ dependence we found that the saturation magnetization
increases with the electron concentration, reaching for $x=$0.1, $M_{0}\sim $1$%
\mu _{B}$/Mn-atom, that is far from the value expected for full FM
alignement, $M_{0}$=(3+$x$)$\mu _{B}$. We have interpreted these results in
terms of magnetic polarons, and we propose that the response of $M$ vs. $H$
is due to the displacement of the $e_{g}$ electrons from one magnetic
sublattice to the other, causing the alignement of the polarons with the
external field (Fig. 7).

\begin{figure}[t]
\centerline{\psfig{figure=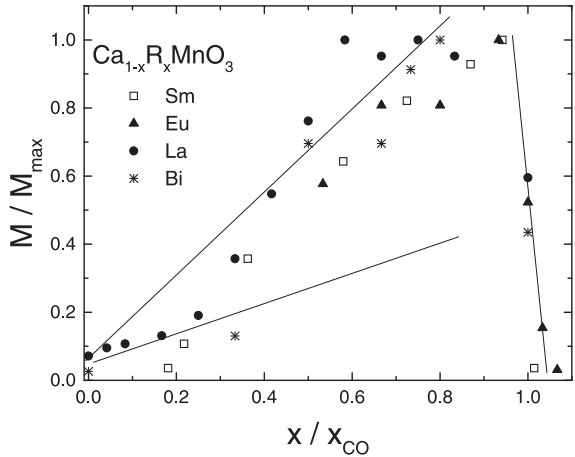,height=6cm,angle=0}}
\medskip
\caption{$M_{0}$ vs. $x$ dependence for Ca$_{1-x}$R$_{x}$MnO$_{3}$ for
different nonmagnetic R dopants. $M_{0}$ and $x$ are normalized to the $M_{max}
$ and $x_{CO}$ values as explained in the text.}
\end{figure}

Our measurements in Ca$_{1-x}$Y$_{x}$MnO$_{3}$ show that the magnetic and
electrical transport properties share a common behaviour with other Ca$_{1-x}
$R$_{x}$MnO$_{3}$ with R$=$Lanthanides, in spite of the important
mismatch caused by the small size of Y$^{3+}$ ions. To make this point
clear, in Fig. 10 we show M(5K) vs. $x$, measured for non magnetic (or weakly
magnetic) R$^{3+}$ in Refs. 4, 10, 11, 13, and 18. Data are normalized to
the charge ordering concentration $x_{CO},$ and to the maximum value
of $M(x)$. This behaviour is coincident with that of Fig. 7 for R=Y: the
magnetization rises slowly for small $x$ and, for $x$/$x_{CO}\simeq$0.25, $M$
increases at larger rate. The net magnetization reached depends
on the size of the polarons and of the electron concentration. The dependence 
found suggests a change of behaviour at $x\sim$0.03.
This result confirms that among the three chemical factors governing the CMR
properties in doped perovskites: carrier density, average size, and size
mismatch, the electron concentration is predominant\cite{Rav} in electron
doped materials. The physical mechanism proposed
here can also be applied to the $M_{0}(x)$ behaviour in other electron doped
manganites with different size of R.

The differential susceptibility , slope of the linear dependence of $M$ vs. $%
H$ in the high field region, results also proportional to the dopant
concentration and is, in all cases, larger than the expected value for
the CaMnO$_{3}$ background. We can interpret the $x$ dependence of $\chi
_{diff}$ as a process of enlargement of the polarons with the applied field
as suggested by numerical simulations\cite{Aliagen}.

In the paramagnetic region we have found that $\chi ^{-1}$($T$) curves
deviate substantially from the Curie-Weiss law, showing strong ferromagnetic
DE correlations, compiting with the AF interaction of the background
material. We discuss our results through a mean field model where FM\ and AF
interactions between Mn$^{4+}$ and Mn$^{3+}$ are included. From the fits of $%
\chi ^{-1}$($T$) and using the measured $T_{mo}$ we obtain values of the
parameters of the model, which indicate that the FM-DE interaction is about
twice as intense as the AF interaction. The effect of the FM coupling
enhances the magnetic moment of the Mn$^{3+}$ ions reflecting the magnetic
polaron formation, even in the PM phase.

The Ca$_{1-x}$Y$_{x}$MnO$_{3}$ conductivity is proportional to the dopant
concentration as shown in Fig. 8(b). This dependence precludes the existence
of phase separation that would produce a percolative behaviour at a critical
concentration. On the other hand, the temperature dependence, for $T>T_{mo}$%
, is of the form $\sigma $($T$)=$T^{-1}$exp($\Delta $/k$_{B}T$). This
behaviour is compatible with polaronic conductivity.

Summing up, from the magnetic and electric transport propeties, we conclude
that the electron doped CaMnO$_{3}$ does not show neither phase separation
nor homogeneus canted structure. Rather, the results point to the scenario
where magnetic or magnetoelastic polarons are responsible for the observed
behaviours.

$Acknowledgements:$ We acknowledge partial support from ANPCYT
(Argentina)-PICT 3-52-1027/3-05266, and CONICET(Argentina)-PIP 4947/96. H. A. is
CONICET (Argentina) PhD-fellow.

\end{document}